\documentclass[usenatbib]{mn2e}

\usepackage{graphicx}
\usepackage{footmisc}
\usepackage{amssymb}

\newcommand{\xmm}{\textit{XMM-Newton }}
\newcommand{\gmm}{$\Gamma$ }  
\newcommand{\nh}{$N\sb{\rm H}$ }                     
\newcommand{\xspec}{\textsc{xspec} }
\newcommand{\chisq}{$\chi\sp{2}$ }
\newcommand{\redchi}{$\chi\sp{2}/\nu$ }
\newcommand{\scott}{S11 }

\title[Detectability of spectral components]{Detectability of low energy X-ray spectral components in type 1 AGN}
\author[A. E. Scott, G. C. Stewart \& S. Mateos]{A. E. Scott$^{1}$\thanks{E-mail: aes25@star.le.ac.uk}, G. C. Stewart$^{1}$ and S. Mateos$^{1,2}$\\
$^{1}$Department of Physics and Astronomy, University of Leicester, University Road, Leicester LE1 7RH, UK\\
$^{2}$Instituto de Fisica de Cantabria (CSIC-UC), Avenida de los Castros, 39005 Santander, Spain }

\begin{document}

\date{Accepted: April 5$^{\rm th}$ 2012.}

\pagerange{\pageref{firstpage}--\pageref{lastpage}} \pubyear{2012}

\maketitle

\label{firstpage}


\begin{abstract}
In this paper we examine the percentage of type 1 AGN which require
the inclusion of a soft excess component and/or significant cold
absorption in the modelling of their X-ray spectra obtained by
\textit{XMM-Newton}.  We do this by simulating spectra which mimic
typical spectral shapes in order to find the maximum detectability
expected at different count levels.  We then apply a correction to the
observed percentages found for the \citet{scott11} sample of 761
sources.  We estimate the true percentage of AGN with a soft excess
component to be $75\pm23\%$, suggesting that soft excesses are
ubiquitous in the X-ray spectra of type 1 AGN.  By carrying out joint
fits on groups of low count spectra in narrow $z$ bins in which
additional spectral components were not originally detected, we show
that the soft excess feature is recovered with a mean temperature kT
and blackbody to power-law normalisation ratio consistent with those
of components detected in individual high count spectra.  Cold
absorption with \nh values broadly consistent with those reported in
individual spectra are also recovered.  We suggest such intrinsic cold
absorption is found in a minimum of $\sim5\%$ of type 1 AGN and may be
present in up to $\sim 10$\%.
\end{abstract}


\begin{keywords}
galaxies: active -- quasars: general -- X-rays: galaxies
\end{keywords}


\section{Introduction}
The X-ray spectral properties of AGN classified optically as type 1
have been recently extensively studied (e.g. \citealt{young09},
\citealt{mateos10}, \citealt{corral11}, \citealt{scott11}).  The
underlying spectrum in the 0.5--12.0 keV band consists of a power law
of photon index \gmm $\sim2$ thought to be produced by the inverse
Compton scattering of low energy disc photons by a corona of
relativistic electrons \citep{haardt93}.  In higher quality spectra, a
soft excess component is detected, rising above the power law at
rest-frame energies $\lesssim$1 keV \citep{arnaud85}.  This was
originally interpreted as the hard tail of the `Big Blue Bump',
accretion disc emission seen in the UV, but is now thought perhaps to
be an artifact of ionised absorption \citep{gierlinskidone04} or
ionised reflection \citep{ross05,crummy06}.  Lower energies are also
affected by photoelectric absorption, although the standard
orientation based Unified Model \citep{antonucci93} does not predict
any intrinsic X-ray absorption to be present in objects which have
been classified as type 1 due to the presence of broad lines in their
UV/optical spectra.

There have been many studies in which some type 1 objects have shown
evidence for intrinsic X-ray absorption.  The typical percentage of
such objects is $\sim 10\%$, with many studies quoting this similar
value \citep{page03, perola04, piconcelli05, mateos05b, mainieri07,
garcet07, young09, mateos10, corral11}.

There are a large range of values quoted for the percentage of type 1
sources which include a soft excess.  The earliest study with
\textit{EXOSAT} suggested 30--50\% of objects included the component
\citep{turner89} and a study with \textit{ASCA} found $\sim 40\%$
\citep{reeves00}.  It has also been suggested that the soft excess may
be a ubiquitous feature in optically selected PG quasars
\citep{porquet04, piconcelli05}, however these samples are biased
towards bright and low redshift sources.  The quoted percentage of
soft excesses can be very different depending upon the redshift range
being considered.  For example, \citet{mateos10} find a percentage of
only 8\% when considering their entire sample, but this is increased
to 36\% when only sources at $z<0.5$ are considered.  Similarly the
CAIXA sample of \xmm target sources finds a high percentage of $\sim
80\%$ \citep{bianchi09}. This could be because the sample is biased
towards low redshift objects and/or good quality spectra in which
detecting the spectral component is easier.  Clearly, in order to
determine whether the soft excess is present in all sources, the
influences of redshift and spectral quality need to be taken into
account, using a sample which covers a large range in these
properties.

In a recent study of the X-ray spectral properties of 761 type 1 AGN,
\citeauthor{scott11} (\citeyear{scott11}; hereafter S11), find $\sim
8\%$ of their sample require a soft excess component and $\sim 4\%$
require intrinsic cold absorption in the modelling of their X-ray
spectra.  It was noted that these values represent lower limits on the
intrinsic percentage of sources which include such components, since
their detectability is limited by the quality of the spectra.  In this
paper we follow on from this analysis and aim to deduce how common
these spectral features really are.  In the case of the soft excess we
do this by simulating typical spectra at different count levels in
order to determine the maximum detection rate expected for such an
additional spectral feature.  This can then be compared to the
observed results in order to determine the intrinsic percentage.  The
original data sample is described in $\S$\ref{section:data}.  The soft
excess simulations and a joint fitting of multiple low count spectra
in an attempt to recover the soft excess are described in
$\S$\ref{section:se}.  Section~\ref{section:abs} considers the
presence of absorption components.  We discuss our results in
$\S$\ref{section:discussion} and summarise our conclusions in
$\S$\ref{section:conclusions}.


\section{Data}
\label{section:data}
\begin{figure}
  \centering
  \begin{tabular}{cc}
    \includegraphics[width=0.22\textwidth]{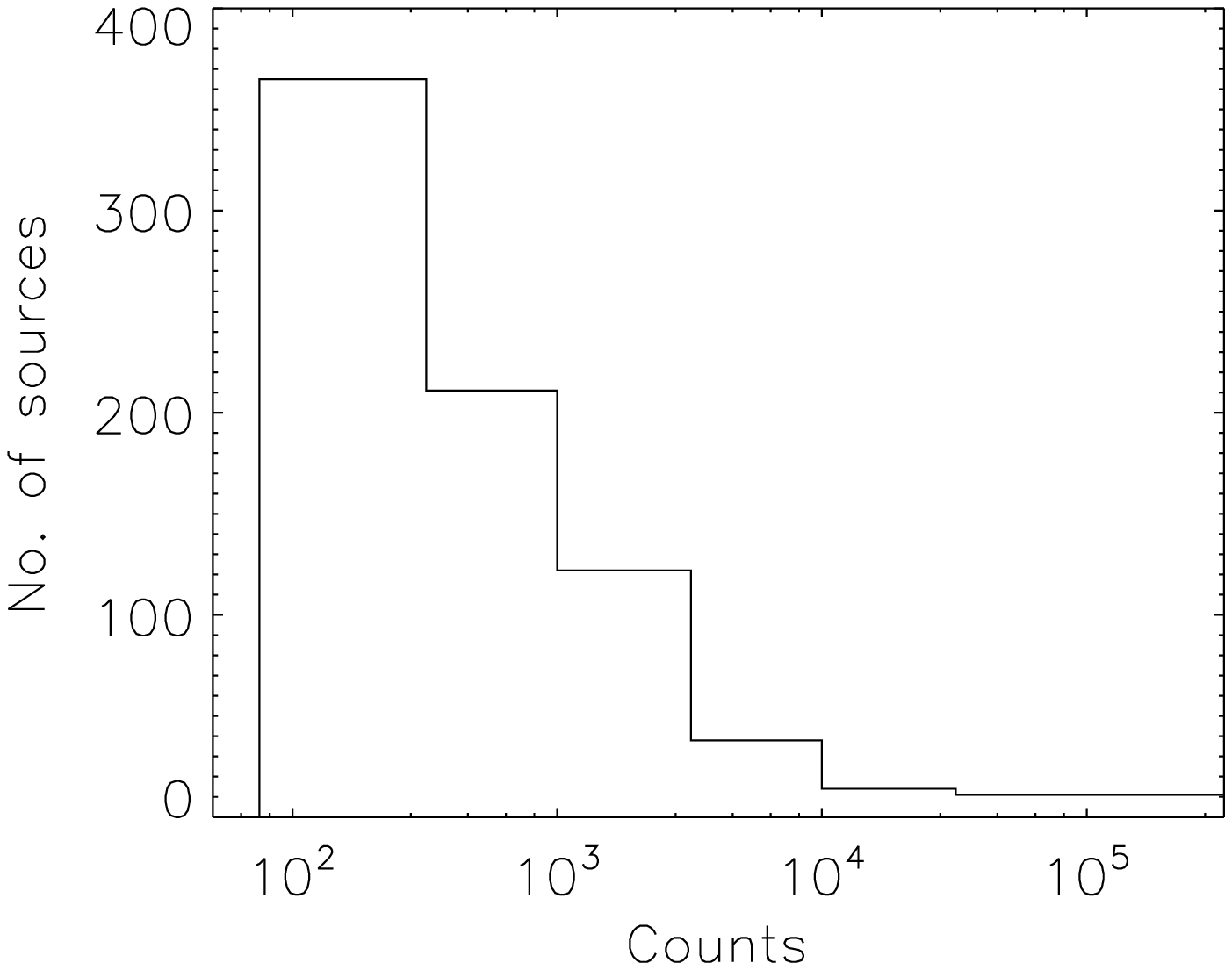} &
    \includegraphics[width=0.22\textwidth]{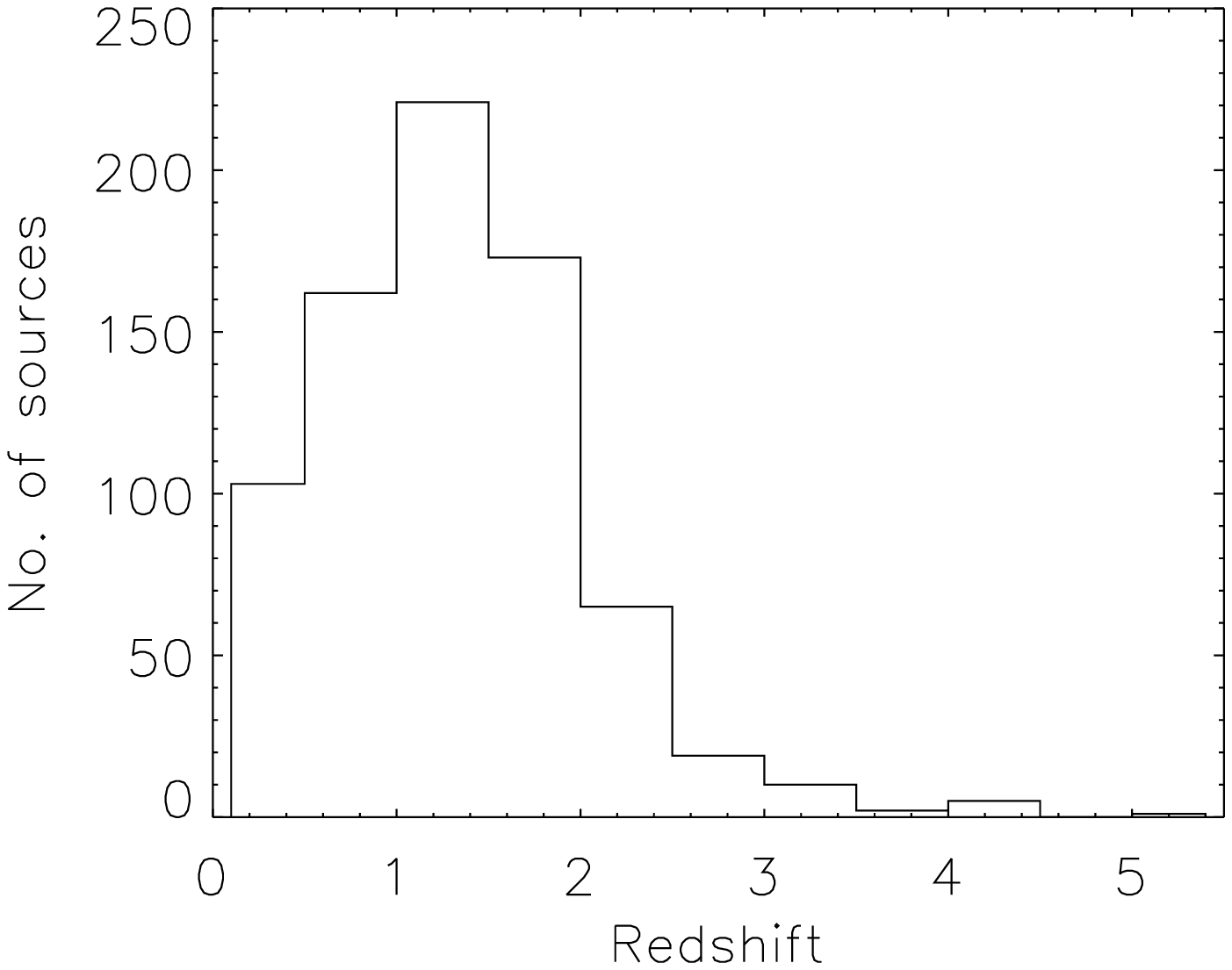}
  \end{tabular}
  \caption{Histograms showing the net counts and redshift distributions of
  the sample used here, and described further in \citet{scott11}.}
  \label{fig:distributions}
\end{figure}

The \scott sample was created by cross-correlating the SDSS DR5 quasar
catalogue \citep{DR5QSO} and the serendipitous X-ray source catalogue,
2XMMi \citep{2XMM}.  The X-ray spectra were extracted using standard
SAS\footnote{The description and documentation are available online at
http://xmm.esac.esa.int/sas/} tasks and fit using \xspec v11.3.2
\citep{xspec} over the energy range 0.5--12.0 keV.  Histograms showing
the redshift and net (i.e. background subtracted) counts distributions
of the sources can be found in Fig.~\ref{fig:distributions}.  All
sources have $>$75 combined MOS+pn counts, allowing both a simple
power law and an absorbed power law to be fit, with the power-law
slope, $\Gamma$, allowed to vary freely.  For 680 sources with $>$100
counts, models including a blackbody component were also considered in
order to model any soft excess.  The best-fitting model was assumed to
be a simple power law unless the F-test, used at the 99\% significance
level, determined that additional components were statistically
required.  A summary of the different models considered, can be found
in Table~\ref{table:models} along with the number of sources best-fit
with each.  Soft excesses are found in $\sim8\%$ of the sources in the
sample and $\sim4\%$ require intrinsic absorption.  An additional
Galactic absorption component was included in each model, fixed at the
\nh value determined from the H\textsc{i} map of \citet{dickey90}.

\begin{table}
\begin{minipage}{75mm}
\centering
  \caption{The different models used to fit the X-ray spectra of the sources in the \scott sample.}
  \label{table:models}
     \centering
     \begin{tabular}{llc}
     \hline
     Model       & Spectral Components               & Sources\\
     \hline
     po          & Power Law                         & 672 \\
     apo         & PL + absorption                   & 29  \\
     po+bb       & PL + soft excess                  & 55  \\
     apo+bb      & PL + absorption + soft excess     & 5   \\
     \hline
                 &                                   & 761 \\
     \hline
  \end{tabular}
\end{minipage}
\end{table}

\begin{figure}
  \centering
  \includegraphics[width=0.5\textwidth]{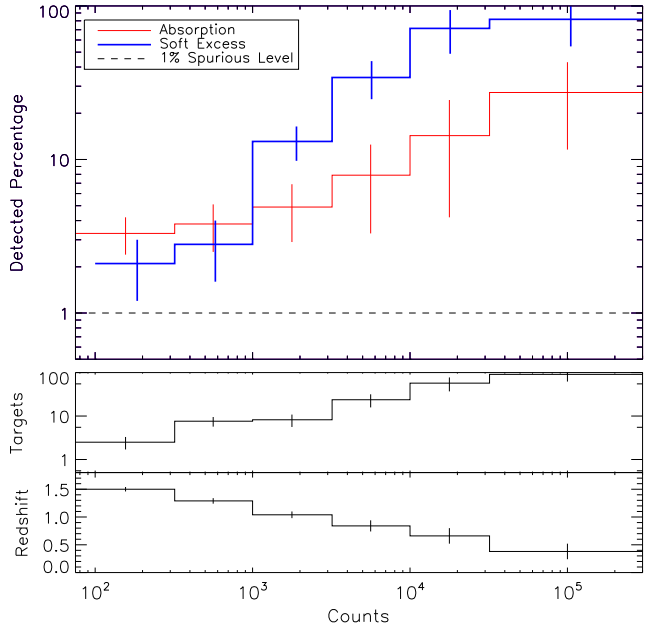}  
  \caption{The top plot shows how the percentage of sources which
  require an absorption or soft excess component varies depending on
  the number of counts (MOS+pn) that are available in the spectra.
  The lowest count bin includes sources with between 75--320 total
  counts for absorption and 100--320 total counts for the soft excess,
  since these were the minimum numbers of counts required for fitting
  that particular spectral component.  The sample includes 62 sources
  which were the target of an \xmm observation, rather than a
  serendipitous detection.  The middle plot shows the percentage of
  sources within each count bin which are target sources.  The errors
  in the top two plots have been calculated from Poissonian
  statistics.  The bottom plot shows how the average redshift of the
  sources in each count bin varies and includes standard errors on the
  mean.}
  \label{fig:detection}
\end{figure}

Fig.~\ref{fig:detection} shows the detected percentages of the soft
excess (thick, blue) and absorption (red) components as a function of
the number of counts in the spectra.  The detected percentage of the
additional components is much lower in spectra with low counts where
the statistics are poorer and the features are not detected with
enough significance.  It was suggested in \scott that since at the
highest count levels we might expect to be able to detect all soft
excess components if they are present, the intrinsic percentage could
be as high as the $\sim 80\%$ found in the highest count bin, making
soft excesses common in the X-ray spectra of type 1 AGN.
Fig.~\ref{fig:detection} also shows intrinsic absorption detected in
up to $\sim 25\%$ of sources in the higher count bins.  Since the
sample is drawn from a population of type 1 AGN, such a component is
not expected to be required in the modelling of their X-ray spectra.
The F-test was used at 99\% significance when choosing the
best-fitting model for a particular source, therefore 1\% of the
detections of a specific spectral component can be considered
spurious.  This 1\% level is shown by the dashed line in
Fig.~\ref{fig:detection}.

The sample contains 62 sources which were the target of an \xmm
observation and therefore generally contain more counts in their X-ray
spectra than the serendipitously detected sources.  The percentage of
targets in each count bin is shown in Fig.~\ref{fig:detection} and
increases towards the higher count bins as expected, with the top bin
including almost only target sources.  These sources could bias the
detection percentages if they were selected for observation due to a
previously known soft excess or intrinsic absorption.  We therefore
exclude the target sources from our subsequent analysis, leaving 699
sources in the sample\footnote{699 sources have $>75$ counts and 619
have $>100$ counts which we fit with models including a soft excess.}.

For sources at increasing redshifts, the contribution of a soft excess
component or absorption in the spectra will gradually decrease as a
larger contribution is redshifted outside of the \xmm EPIC instrument
bandpass \citep{mos,pn}.  Therefore the detected percentages of these
components are expected to be higher in bins containing mostly low
redshift sources, which tend to be the bins with higher numbers of
counts, hence the higher detection rate may be due to this redshift
bias.  The average redshift of the sources in each count bin is
plotted in Fig.~\ref{fig:detection} and does decrease with increasing
counts as expected.  However we note that within each count bin the
sources do cover a large range in redshifts.
  
\begin{figure}
  \centering
  \includegraphics[width=0.45\textwidth]{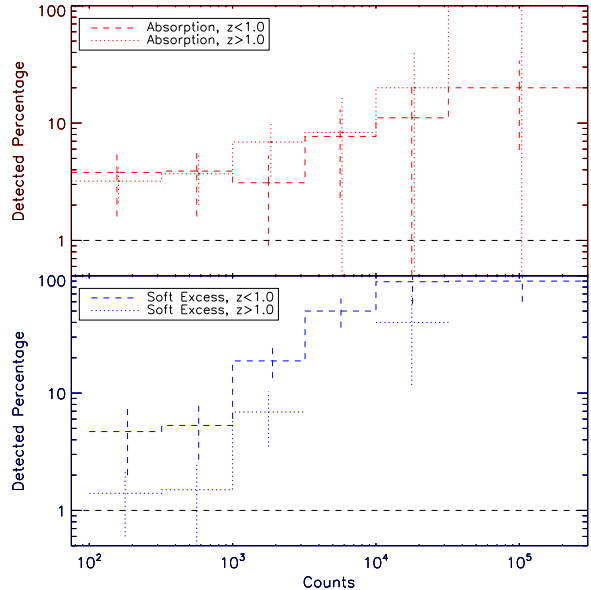}  
  \caption{These figures show how the detected fraction of absorption
  (top) and soft excess (bottom) components varies with the numbers of
  counts in the spectra, considered in broad redshift bins of $z<1$
  and $z>1$.  There is a notable difference between the two curves in
  the bottom panel indicating a strong redshift dependence on the
  detectability of soft excess components.}
  \label{fig:detection_z}
\end{figure}

To further investigate the redshift issue, the sources are split into
broad redshift bins of $z<1$ and $z>1$.  Separate detection curves are
created and are shown in Fig.~\ref{fig:detection_z}.  In the case of
the absorbed sources (top), the detected percentages for the low and
high redshift sources appear to be consistent within the error ranges.
This is likely due to the large range of \nh values found in the
sample, which means that we are able to detect absorption in sources
at a range of redshifts.  The percentage of sources detected with a
soft excess component is higher in the low redshift sample than in the
high redshift sample as expected.  The curve for the $z>1$ sources
shows gaps where the detected fraction falls to zero, due to the low
numbers of sources; although soft excesses are detected in the sample
up to $z=2$, the majority (82\%) are found in sources with $z\lesssim
1$ as expected.

\section{Soft excess components}
\label{section:se}

\subsection{Intrinsic Percentage}
\label{section:sim_se}
In order to determine the intrinsic percentage of sources with a soft
excess component, we carry out multiple sets of simulations in order
to quantify their detectability.  This is done by finding the maximum
percentage of components that are expected to be detected in spectra
with different numbers of counts.

At each of 5 redshifts and between 7 and 11 different count levels, we
simulate 1000 spectra which include a soft excess.  Each spectrum is
fit over the energy range 0.5--12.0 keV with the `po' and `po+bb'
models, and the F-test is used at 99\% significance to determine
whether the component is statistically required.  The percentage of
sources in which we significantly detect the component is then
determined.  By repeating this procedure with sets of spectra at
different count levels and redshifts we construct synthetic
detectability curves from which we can determine the maximum detection
percentage at any count level.

We create our spectra, using the \texttt{fakeit} command in \xspec
which distributes a given number of counts, controlled by varying the
exposure time, around a defined model with statistical fluctuations
and assigns them Poissonian errors.  We define the model such that it
mimics the shape of a typical source in the \scott sample, both in
terms of the shape of the components i.e. the \gmm and kT values and
the size of the components, particularly the ratio of the blackbody
normalisation to that of the power-law normalisation since this will
also determine how easy the blackbody is to detect over the power-law
continuum.  55 sources required the `po+bb' model in the original
\scott analysis.  The distribution of power-law slopes and kT values
for these sources were each fit with a Gaussian, and the best-fitting
mean values with dispersions were \gmm$=1.79\pm0.46$ and
$\textrm{kT}=0.17\pm0.08 \textrm{ keV}$.  We therefore simulate
sources with \gmm$=1.8$, $\textrm{kT}=0.2 \textrm{ keV}$ and fix the
normalisation ratio to the median value of 0.04.  These values are
intended to represent a `typical' source rather than the full range of
values, although the distributions of kT and normalisation ratio are
reasonably narrow.  As shown in fig. 22 of S11, there is a very tight
correlation between the luminosities of the blackbody and power-law
components.  The median normalisation ratio of 0.04 used here
corresponds to a luminosity ratio of $\sim0.2$.

\begin{figure}
  \centering
  \includegraphics[width=0.45\textwidth]{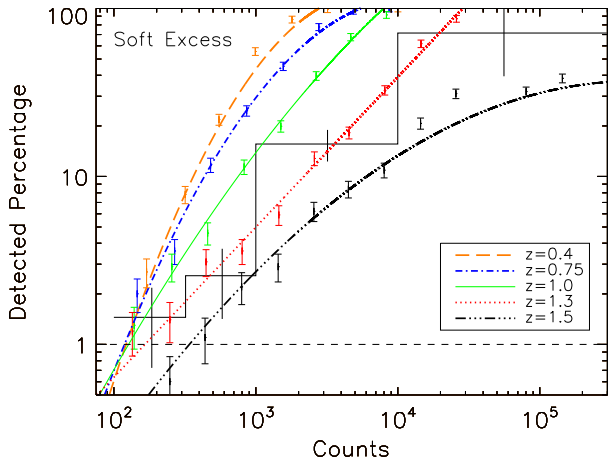}
  \caption{The solid black line with error bars shows the observed
  percentage of soft excess components as a function of counts for the
  \scott sample after the target sources have been removed, which
  required a rebinning of the sources.  The coloured curves show the
  detectability level for a soft excess component with a typical shape
  and size.  Sets of simulations are carried out at different
  redshifts and count levels as shown by the coloured error bars. The
  curves are created from a quadratic fit to at least 7 such results
  (a linear fit is used for $z=1.3$).}
  \label{fig:detection_curve}
\end{figure}

As has been previously discussed, the detectability of soft excess
components is strongly dependent upon the redshift of the source.
Therefore we run sets of simulations at a range of redshifts ($z=0.4$,
$z=0.75$, $z=1.0$, $z=1.3$ and $z=1.5$) and create five separate
detectability curves shown by the different lines in
Fig.~\ref{fig:detection_curve}.  For interpolation purposes they are
fit with a quadratic function which describes four of the curves well.
The curve for $z=1.3$ has a slightly different curvature from the
asymptotic behaviour expected and therefore in this case we use a
linear function to fit the simulated points.  The curves indicate the
detection percentage that we would expect to see for a particular
count level if all sources contained this component.  If the observed
percentages were consistent with this line, it would indicate that the
spectral features are present in all of the sources in our sample and
it is merely the quality of the spectra that limits our ability to
detect them.  For sources at $z\leq1$ with $\gtrsim10000$ counts, the
detectability curves lie at 100\%.  This suggests that the intrinsic
percentage could be as high as $\sim70\%$; the percentage found in the
highest count bin (after removal of the target sources).  We can
determine the true soft excess detection percentage (corrected
percentage) for each count bin in our data by dividing the observed
detection percentage by the maximum percentage obtained from our
simulated curves which is fixed to lie between the spurious level,
$1\%$, and $100\%$.

As each bin includes sources at a range of redshifts, we calculate a
corrected percentage using all five of the detectability curves and
determine a weighted average value according to the equation;
\begin{equation}
\textrm{Result} = \sum{\left(\textrm{weight} \times \textrm{corrected percentage}\right)}
\label{eqn:weight}
\end{equation}
where the `weight' is the fraction of sources for which each
particular redshift curve is appropriate for within that count bin.
For sources at $z>2$, the maximum percentage is fixed at 1\%, since no
real soft excesses are expected, but spurious detections may occur.
Since the count bins are broad and the sources are not distributed
evenly within them, they are further divided into sub-bins.  The
redshift-corrected percentage is calculated for each as outlined
above, and an overall corrected percentage is reconstructed for the
full count bin using equation~\ref{eqn:weight}, where `weight' in this
case is the fraction of sources within the particular count sub-bin.

\begin{figure}
  \centering
  \includegraphics[width=0.45\textwidth]{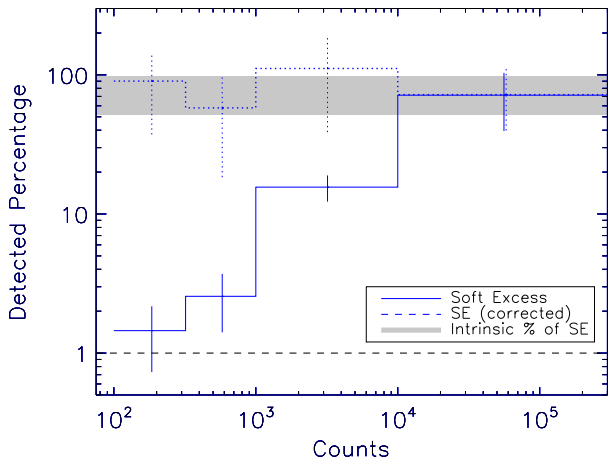}
  \caption{The blue solid line shows the observed percentage
  of soft excess components in the \scott sample (after the target
  sources have been removed).  The dashed line shows the detected
  percentage determined after the correction for spectral quality.
  The grey bar indicates the location of the intrinsic percentage of
  sources with soft excesses, where the width of the bar represents
  the 1$\sigma$ error boundary.}
  \label{fig:corrected_fig}
\end{figure}

The dashed blue line in Fig.~\ref{fig:corrected_fig} shows the
corrected percentages\footnote{We note that percentages $>100\%$ are a
possible consequence of using this correction factor method.  The
count bins have been re-defined in order to reduce this statistical
effect.  However, should the intrinsic percentage be exactly 100\%, we
would expect some bins to show correction percentages lower, and
others greater than 100\%.}.  The values are roughly constant after
the effect of spectral quality has been removed.  In order to
determine a value for the intrinsic percentage, \chisq is calculated
for constant percentages between 0 and 100\%.  The minimum \chisq
occurs for a constant percentage of $75\pm23\%$, for which the fit has
a null hypothesis probability of $p=91\%$.  This method produces a
corrected percentage greater than 100\% in the 3rd bin.  Whilst this
has no physical meaning, its large error bar makes it entirely
consistent with the corrected percentages obtained for the other bins.
Capping this bin at 100\% reduces our intrinsic percentage estimate by
only 1\%.

To test whether a 1\% spurious level is appropriate we simulated
spectra which did not include a soft excess, and determined in how
many we incorrectly detected this component with $>99\%$ significance.
For the majority of count levels this percentage was consistent with,
or lower than 1\%. However, for the lowest and highest count levels
($\lesssim 100$ and $\sim 50,\!000$) the spurious level was somewhat
higher (up to $\sim5\%$), suggesting that both the usual statistical
problems affecting simulated fits and some small systematic
calibration errors exist.  The count bin most affected by the spurious
level is the lowest, as the simulated curve for $z=1.5$ sources falls
below 1\%.  If the maximum percentage from the simulated curve is
fixed at 1\%, the spurious level, the correction factor is
under-estimated.  Excluding this bin from the intrinsic percentage
determination gives $71\sp{+26}\sb{-24}\%$, which is consistent with
the value obtained when the bin is included.

\subsection{Joint Spectral Fitting}
\label{section:joint_se}
We have shown that after correcting for spectral quality, a soft
excess component may be ubiquitous in the \scott sample.  We now
re-analyse spectra with low numbers of counts in a joint fitting to
see if the soft excess feature can be recovered in a combination of
spectra where it was previously undetected.

The \scott sample includes 436 sources which were originally fit with
a simple power-law model i.e. the soft excess feature was not detected
in their spectra with $>99\%$ significance, and which have $<500$
counts.  This limit is imposed so that the joint fitting is not
dominated by a single object with high numbers of counts.  Samples of
$\sim30$ sources ($\sim50$ MOS and/or pn spectra) are created which
cover a narrow range in redshift.  The resulting groups of spectra
include a total of $\sim 7000$ counts, a level at which a soft excess
in a $z<1$ source is expected to be detected 100\% of the time if it
is present.  This is shown in Fig.~\ref{fig:detection_curve} where the
detectability curves for $z=0.4$ and $z=0.75$ sources are at $100\%$
at a $7000$ count level.

The groups of spectra are jointly fit with the simple power law model,
`po', in which \gmm is free to vary.  In each case we find a
best-fitting result of \gmm$\sim2$.  In the case of the lowest two
redshift bins the fit is significantly improved by using the `po+bb'
model, both in terms of a lower $\chi\sp{2}/\nu$, as listed in
Table~\ref{table:joint_bb}, and an F-test comparing the two models
which equals 100\% in both cases.  However the \gmm values are lower
than 1.8, the value used in the simulations of the previous section.
At lower redshifts, we expect to be more sensitive to cold absorption
(see $\S$\ref{section:abs}) and we note that if a fixed absorption
component of \nh$=10\sp{21} \textrm{ cm}\sp{-2}$ is included in the
fit, steeper values of \gmm$\sim 1.70\pm0.05$ are obtained.  This
does not fully correct for the effect as the resulting continuum from
stacking spectra with different levels of absorption is known to be
flatter \citep{mateos10}.

We also fit each group of spectra with the `po+bb' model in which \gmm
is fixed at 1.8 and kT is tied to the same value for each of the
spectra. We determine a single, best-fitting normalisation ratio
common to all the spectra by fixing the power-law normalisations at
$10\sp{-5}$ and tying the blackbody normalisations for each spectrum
to a common value.  A freely varying constant is added to the model to
allow each data set to vary independently.  The results can be found
in Table~\ref{table:joint_bb} listed as the model `po+bb fix A'.  For
all except the 1st bin we find average normalisation ratios which are
consistent with the value of 0.04 used in the simulations of the
previous section.  For all except the 5th bin, the kT values are
consistent with those found in individual sources with high count
spectra.

Each group of spectra is also fit with the `po+bb' model in which \gmm
is fixed at 1.8, kT is tied to a common value for each of the spectra
and the normalisations of both the blackbody and power-law components
are left free to vary.  These results are listed in
Table~\ref{table:joint_bb} as the model `po+bb fix B', and include the
median value of the blackbody to power-law normalisation ratios and
the percentage of spectra for which the ratio is $>0.01$, which is
interpreted as a blackbody component being present.  For the three
lower redshift bins ($z\lesssim1$) the inclusion of a blackbody
component in the spectral fit results in a better fit and the
blackbody temperatures are consistent with the previous values
obtained from the individual high count spectra.  The normalisation
ratios of the blackbody to power-law components are also consistent,
both in terms of the median value and a Kolmogorov--Smirnov (KS) test
which finds the full distribution of ratios to be not significantly
different from those found in single object spectra in S11.  The
median normalisation ratio of $\sim0.04$ is consistent with the value
used in the simulations of the previous section and corresponds to a
luminosity ratio of $\sim0.2$.  In addition, the percentage of spectra
which do include a blackbody component (since the fit allows a
blackbody normalisation of zero for individual spectra) is consistent
with the intrinsic percentage calculated in $\S$\ref{section:sim_se}.
This model also gives a better fit for the highest of the redshift
bins.  However, the kT value is too high to be consistent with the
temperatures observed in individual spectra.

\begin{table*}
\centering
\begin{tabular}{lccllllcc}
\hline
\multicolumn{1}{l}{$z$ range}&
\multicolumn{1}{c}{No. of sources}&
\multicolumn{1}{c}{Total} &
\multicolumn{1}{l}{Model} &
\multicolumn{1}{l}{\redchi} &
\multicolumn{1}{l}{\gmm} &
\multicolumn{1}{l}{kT} &
\multicolumn{1}{c}{Normalisation} &
\multicolumn{1}{c}{Percentage} \\
\multicolumn{1}{l}{}&
\multicolumn{1}{c}{(spectra)$^a$}&
\multicolumn{1}{l}{Counts}&
\multicolumn{1}{l}{}&
\multicolumn{1}{l}{}&
\multicolumn{1}{l}{}&
\multicolumn{1}{l}{(keV)}&
\multicolumn{1}{c}{Ratio}&
\multicolumn{1}{c}{with SE}\\
\hline
0.21--0.55 & 34 (51) & 8078 & po          & 859/636 (1.351) & $2.10\sp{+0.04}\sb{-0.05}$ &                               &                   &             \\[0.1pc]
           &         &      & po+bb       & 664/584 (1.137) & $1.58\sp{+0.08}\sb{-0.06}$ & $0.23\sp{+0.01}\sb{-0.03}$    & 0.07              & $78 \pm 17$ \\[0.1pc]
           &         &      & po+bb fix A & 873/635 (1.375) & 1.8 (fixed)                & $0.23\sp{+0.03}\sb{-0.03}$    & $0.024 \pm 0.002$ &             \\[0.1pc]
           &         &      & po+bb fix B & 675/585 (1.154) & 1.8 (fixed)                & $0.19\sp{+0.02}\sb{-0.02}$    & 0.04              & $74 \pm 16$ \\[0.5pc]
0.56--0.77 & 30 (50) & 7661 & po          & 861/615 (1.400) & $2.07\sp{+0.05}\sb{-0.05}$ &                               &                   &             \\[0.1pc]
           &         &      & po+bb       & 682/564 (1.209) & $1.62\sp{+0.08}\sb{-0.13}$ & $0.24\sp{+0.03}\sb{-0.02}$    & 0.07              & $84 \pm 18$ \\[0.1pc]
           &         &      & po+bb fix A & 854/614 (1.391) & 1.8 (fixed)                & $0.19\sp{+0.04}\sb{-0.04}$    & $0.040 \pm 0.010$ &             \\[0.1pc]
           &         &      & po+bb fix B & 693/565 (1.227) & 1.8 (fixed)                & $0.20\sp{+0.03}\sb{-0.02}$    & 0.05              & $72 \pm 16$ \\[0.5pc]
0.77--0.98 & 33 (51) & 7367 & po          & 683/611 (1.118) & $2.03\sp{+0.05}\sb{-0.04}$ &                               &                   &             \\[0.1pc]
           &         &      & po+bb       & 624/559 (1.116) & $1.98\sp{+0.04}\sb{-0.05}$ & $0.044\sp{+0.002}\sb{-0.001}$ &                   &             \\[0.1pc]
           &         &      & po+bb fix A & 681/610 (1.116) & 1.8 (fixed)                & $0.27\sp{+0.05}\sb{-0.05}$    & $0.036 \pm 0.005$ &             \\[0.1pc]
           &         &      & po+bb fix B & 579/560 (1.034) & 1.8 (fixed)                & $0.28\sp{+0.03}\sb{-0.04}$    & 0.06              & $63 \pm 14$ \\[0.5pc]
0.98--1.12 & 31 (51) & 7583 & po          & 632/629 (1.005) & $2.02\sp{+0.05}\sb{-0.04}$ &                               &                   &             \\[0.1pc]
           &         &      & po+bb       & 576/577 (0.998) & $1.97\sp{+0.04}\sb{-0.06}$ & $0.053\sp{+0.003}\sb{-0.002}$ &                   &             \\[0.1pc]
           &         &      & po+bb fix A & 616/628 (0.981) & 1.8 (fixed)                & $0.26\sp{+0.04}\sb{-0.04}$    & $0.048 \pm 0.008$ &             \\[0.1pc]
           &         &      & po+bb fix B & 629/578 (1.088) & 1.8 (fixed)                & 0.02$^b$                      &                   &             \\[0.5pc]
1.12--1.19 & 30 (50) & 6749 & po          & 604/578 (1.045) & $2.03\sp{+0.04}\sb{-0.05}$ &                               &                   &             \\[0.1pc]
           &         &      & po+bb       & 571/527 (1.083) & $1.98\sp{+0.04}\sb{-0.06}$ & $0.06\sp{+0.01}\sb{-0.03}$    &                   &             \\[0.1pc]
           &         &      & po+bb fix A & 597/577 (1.035) & 1.8 (fixed)                & $0.35\sp{+0.05}\sb{-0.05}$    & $0.045 \pm 0.005$ &             \\[0.1pc]
           &         &      & po+bb fix B & 606/528 (1.148) & 1.8 (fixed)                & $0.065\sp{+0.005}\sb{-0.005}$ &                   &             \\[0.5pc]
1.19--1.28 & 29 (50) & 5614 & po          & 485/509 (0.953) & $1.95\sp{+0.05}\sb{-0.05}$ &                               &                   &             \\[0.1pc]
           &         &      & po+bb       & 450/458 (0.983) & $1.89\sp{+0.05}\sb{-0.06}$ & $0.062\sp{+0.002}\sb{-0.002}$ &                   &             \\[0.1pc]
           &         &      & po+bb fix A & 471/508 (0.927) & 1.8 (fixed)                & $0.24\sp{+0.06}\sb{-0.06}$    & $0.050 \pm 0.030$ &             \\[0.1pc]
           &         &      & po+bb fix B & 416/459 (0.906) & 1.8 (fixed)                & $0.34\sp{+0.07}\sb{-0.06}$    & 0.04              & $66 \pm 15$ \\[0.1pc]
\hline
\end{tabular}
\caption{The samples used for the joint fitting and the results from
fitting the spectra with the simple power law model `po' in which \gmm
is allowed to vary freely and the `po+bb' model in which \gmm is both
fixed at 1.8 and free to vary. `po+bb fix A' refers to the model in
which the blackbody to power-law normalisation ratio is kept the same
for each spectrum and `po+bb fix B' is the model in which both the
blackbody and power-law normalisations are free to vary.  In the case
of model A, the normalisation ratio quoted is the best-fitting value
with a 68\% error and for model B it is median value.  The errors on
the kT and \gmm parameters are 90\%.  Notes: $^a$Not all sources have
both a MOS and a pn spectrum available. $^b$For this sample we are
unable to estimate reliable errors from the fit.}
\label{table:joint_bb}
\end{table*}

It has been suggested that the soft excess feature is ubiquitous in
high accretion rate AGN and that it is this parameter which may
determine its presence in the spectra, or the size of the component
(e.g. \citealt{done11}).  We use the Eddington ratio as a proxy for
mass accretion rate defined as $\lambda\sb{\rm Edd} =
\textrm{log}(L\sb{\rm bol}/L\sb{\rm Edd})$.  The bolometric luminosity
was estimated from the observed X-ray luminosity in the 2--10 keV
band, measured from the best-fitting spectral model, by applying the
luminosity-dependent correction of \citet{marconi04}.  The Eddington
luminosity was estimated using virially determined black hole mass
estimates from the \citet{shen08} catalogue.  As before, we use only
sources with $<500$ total counts and those best-fit with the model
`po'.  In addition, we also further restrict the sources to those at
$z<1$ and we split the remaining sources into 3 sub-samples containing
roughly equal numbers of sources.  The samples are fit with both the
`po+bb fix A' and `po+bb fix B' models, the results of which are
listed in Table~\ref{table:joint_edd} along with the properties of
each sample.  The `po+bb fix A' model finds that a best-fitting
normalisation ratio is consistent with the 0.04 found in single
spectra for the top 2 bins.  In the case of the lowest $\lambda\sb{\rm
Edd}$ bin this value is considerably lower, although it has a large
associated error.  A KS test finds the distributions of the
normalisation ratios determined from the `po+bb fix B' model to be
consistent, however the median values do vary from a lower value than
expected in the low $\lambda\sb{\rm Edd}$ bin to a higher value than
expected in the high $\lambda\sb{\rm Edd}$ bin.  In addition, the
percentage of spectra for which the normalisation ratio is $>0.01$ is
consistent with the intrinsic percentage in the case of the top 2
bins, but is lower, although still within errors, for the lowest
$\lambda\sb{\rm Edd}$ bin.  Whilst the evidence is not strong, this
may suggest that soft excesses are smaller and less common in sources
with lower accretion rates and we do not rule this out.
 
\begin{table*}
\centering
\begin{tabular}{lccllllcc}
\hline
\multicolumn{1}{l}{$\lambda\sb{\rm Edd}$ range$^a$}&
\multicolumn{1}{c}{No. of sources}&
\multicolumn{1}{c}{Total} &
\multicolumn{1}{l}{Model} &
\multicolumn{1}{l}{\redchi} &
\multicolumn{1}{l}{\gmm} &
\multicolumn{1}{l}{kT} &
\multicolumn{1}{c}{Normalisation} &
\multicolumn{1}{c}{Percentage} \\
\multicolumn{1}{l}{}&
\multicolumn{1}{c}{(spectra)}&
\multicolumn{1}{l}{Counts}&
\multicolumn{1}{l}{}&
\multicolumn{1}{l}{}&
\multicolumn{1}{l}{}&
\multicolumn{1}{l}{(keV)}&
\multicolumn{1}{c}{Ratio}&
\multicolumn{1}{c}{with SE}\\
\hline
$-2.3 \textrm{ to} -1.3$ & 28 (47) & 5613 & po+bb fix A & 685/498 (1.376) & 1.8 (fixed) & $0.15\sp{+0.16}\sb{-0.10}$ & $0.008 \pm 0.7$   &            \\[0.1pc]
                         &         &      & po+bb fix B & 570/452 (1.261) & 1.8 (fixed) & $0.18\sp{+0.04}\sb{-0.02}$ & 0.005             & $45 \pm 12$\\[0.5pc]
$-1.3 \textrm{ to} -1.0$ & 32 (52) & 8383 & po+bb fix A & 821/660 (1.244) & 1.8 (fixed) & $0.25\sp{+0.03}\sb{-0.02}$ & $0.048 \pm 0.004$ &            \\[0.1pc]
                         &         &      & po+bb fix B & 657/609 (1.079) & 1.8 (fixed) & $0.23\sp{+0.02}\sb{-0.02}$ & 0.04              & $75 \pm 16$\\[0.5pc]
$-1.0 \textrm{ to} +0.2$ & 26 (36) & 6536 & po+bb fix A & 546/500 (1.092) & 1.8 (fixed) & $0.18\sp{+0.04}\sb{-0.03}$ & $0.05 \pm 0.01$   &            \\[0.1pc]
                         &         &      & po+bb fix B & 472/465 (1.015) & 1.8 (fixed) & $0.16\sp{+0.04}\sb{-0.03}$ & 0.084             & $83 \pm 21$\\[0.1pc]
\hline
\end{tabular}
\caption{The samples created based on Eddington ratio values and the
results of the joint spectral fitting.
$^a\lambda\sb{\rm Edd}=\textrm{log}\left(L\sb{\rm bol}/L\sb{\rm Edd}\right)$}
\label{table:joint_edd}
\end{table*}


\section{Absorption components}
\label{section:abs}
\begin{figure}
  \centering
  \includegraphics[width=0.45\textwidth]{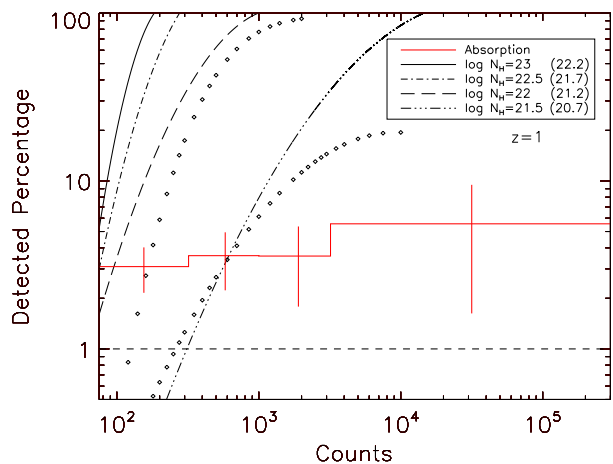}
  \caption{This figure shows how the detected percentage of an
intrinsic cold absorption component varies with the number of counts
in the X-ray spectra (red, solid line).  It reproduces
Fig. ~\ref{fig:detection}, but the sources which were the target of an
\xmm observation have been removed.  As a result of the reduced number
of sources this leaves, the original top three bins are combined.  The
percentage appears to remain constant at $\sim 3\%$ for spectra with
75--32,000 counts, suggesting the detectability is not as heavily
dependent on the spectral quality as is the case for the soft excess.
The black lines represent simulated detectability curves for the
detection of absorption components with different column densities
(shown by the different line styles) in simulated spectra at $z=1$.
The equivalent \nh value at $z=0$ is quoted in brackets on the figure.
The open circles show the detectability of absorption of log \nh $=
22$ and log \nh $= 21.5$, when a soft excess of typical shape and size
is also included in the spectra.}
  \label{fig:absorption}
\end{figure}

Intrinsic cold absorption may also be present in type 1 AGN,
suppressing the lower energy emission.  The percentage of sources with
detected absorption is shown in Fig.~\ref{fig:detection} and appears
to be limited by the spectral quality in a similar way to that of the
soft excess.  However, the effect is not as strong, with the detected
percentage decreasing by approximately $20\%$ from the higher to lower
count bins rather than $\sim80\%$ in the case of the soft excess.  It
was suggested in \scott that the true percentage of absorbed sources
could be as high as the $\sim25\%$ found in the highest count range,
where we might expect the spectra to be of good enough quality to
detect any significant absorption if present.  However, the highest
count bins are heavily contaminated by target sources, resulting in a
lower detected percentage of absorption components once they are
removed from consideration.  Fig.~\ref{fig:absorption} shows the
detected percentage after target removal.  The highest count bin in
this plot now suggests that $5.6 \pm 3.9\%$ of type 1 AGN may include
an intrinsic absorption component.  In the lower count bins ($<3200$),
the detected percentage of absorbed sources does not vary
significantly suggesting that the detectability is not as heavily
dependent on spectral quality as it is for the soft excess feature.

The detectability of absorption in the spectra is highly dependent on
the \nh value and the range of rest-frame column densities found in
the absorbed sources is very broad, $10\sp{21}$ to
$10\sp{23}\textrm{cm}\sp{-2}$.  Therefore we cannot choose a single
model to simulate which is representative of all the absorbed sources
we detect, unlike in the case of the soft excess, and we do not have
enough statistics in order to weight detectability curves by both \nh
and $z$.  We do attempt to quantify the detectability of absorption
components with different column densities by simulating absorbed
spectra at $z=1$.  These detectability curves are shown in
Fig.~\ref{fig:absorption} by the black curves of different line style.
They show that a column of $10\sp{23} \textrm{cm}\sp{-2}$ (equivalent
to $2\times10\sp{22}\textrm{cm}\sp{-2}$ at $z=0$) would be detected in
most spectra with $>200$ counts, whereas a column of $\sim 3 \times
10\sp{21} \textrm{cm}\sp{-2}$ (equivalent to
$5\times10\sp{20}\textrm{cm}\sp{-2}$ at $z=0$) is not strong enough to
be detected in spectra of this quality at $z=1$.  For the highest
count bin ($\gtrsim3200$ counts), the detectability curves for all but
the lowest level of \nh shown in Fig.~\ref{fig:absorption} are at
100\%.  This means that we expect to be sensitive to all reasonable
levels of \nh and therefore our intrinsic percentage estimate is
robust.  The fraction of sources with particular \nh levels is roughly
constant in both different $z$ and count bins (the 2 properties being
correlated), and hence the slight decrease in the detected percentage
between the top bin where we expect to detect all levels of \nh and
the bottom bins is what is expected when objects with lower \nh and/or
higher redshifts are no longer detectable.

These simulations do not include a soft excess component which could
also reduce the detectability of any absorption present.  We
investigate this by simulating sources with both the standard soft
excess parameters and 2 values of \nh (shown by the lines of open
circles in Fig.  \ref{fig:absorption}). For log \nh$=22$ our
sensitivity drops by approximately $10\%$ at low count levels
($\sim200$), increasing to $\sim25\%$ at higher count levels
($\sim1000$), making little difference to our conclusions. In the case
of log \nh$=21.5$, at low count levels ($\lesssim1000$) we are mostly
insensitive to the absorption anyway, such that the inclusion of a
soft excess makes little difference to its detectability.  At higher
counts levels ($\sim10,\!000$) where the statistics are better,
including the soft excess can reduce our sensitivity to the absorption
component by $\sim65\%$.

In $\S$\ref{section:joint_se}, a joint fitting was carried out on
groups of low count spectra to see if the soft excess feature could be
recovered.  Similarly, we fit the `apo+bb' model to the same samples
to see if an absorption component can be recovered in addition to the
soft excess already found to be present.  We implement the model with
\gmm fixed to 1.8, kT fixed at 0.2 keV and leave both the power law
and blackbody normalisations free to vary.  The best-fitting \nh
values are listed in Table~\ref{table:joint_abs}.  We find that an
absorption component can be recovered in the low count spectra, but
only in the lower redshift bins is this component constrained.
Although the \nh values are of the order of those seen in single
object fits in the \scott sample, the range in column densities means
that the values we obtain here merely represent an `average' \nh
value, the exact value of which should be treated with caution.

\begin{table}
\centering
\begin{tabular}{cc}
\hline
\multicolumn{1}{c}{$z$ range}&
\multicolumn{1}{c}{\nh $(\times 10\sp{22} \textrm{cm}\sp{-2})$}\\
\hline
0.21--0.55 & $0.57\sp{+0.27}\sb{-0.15}$\\[0.1pc]
0.56--0.77 & $0.83\sp{+0.22}\sb{-0.19}$\\[0.1pc]
0.77--0.98 & $<0.43$          \\[0.1pc]
0.98--1.12 & $<0.17$          \\[0.1pc]
1.12--1.19 & $<0.53$          \\[0.1pc]
1.19--1.28 & $<0.61$          \\
\hline
\end{tabular}
\caption{Results from fitting the `apo+bb' model to the groups of low
count spectra described in Table~\ref{table:joint_bb}.  In each case,
\gmm and kT are fixed at 1.8 and 0.2 keV respectively.  90\% errors
are quoted on the \nh value.}
\label{table:joint_abs}
\end{table}


\section{Discussion}
\label{section:discussion}
The origin of the soft excess emission is still a matter of debate.
Since this feature has been shown to be ubiquitous in the X-ray
spectra of type 1 AGN, any models to explain its origin must be
applicable to all sources.  Some theories of the soft excess describe
it as an `apparent' feature, rather than an additional component in
the spectrum.  A strong jump in opacity at $\sim 0.7 $ keV is created
by lines and edges of ionised O\textsc{vii} and O\textsc{viii} which
are smeared by the high velocities or gravitational redshifts found
close to a black hole.  This can appear from absorption through
optically thin material in the line of sight \citep{gierlinskidone04}
or via reflection from optically thick material out of the line of
sight \citep{ross05,crummy06}.  More recently, \citet{done11} have
suggested that the soft excess may be intrinsic emission from the
disc, which is shifted into soft X-ray energies due to the required
colour-temperature correction, and further Compton upscattering
produces the full components observed.  However, this only applies to
the lowest mass/highest accretion rate AGN and since we find the soft
excess to be present not only in high accretion rate sources, this
suggests two separate interpretations for the soft excess are
required.  An alternative theory, is that part of the soft X-ray
emission may be due to cooling outflows which are now thought to be
reasonably common in AGN \citep{tombesi10}.  During Eddington
accretion episodes, high velocity ($v \sim 0.1c$) and highly ionised
($\xi \sim 10\sp{4}$) winds are produced and when these interact with
the interstellar medium of the host galaxy, the gas is strongly
shocked \citep{king10}.  Subsequent Compton cooling of this gas may be
observable as a soft X-ray component, as suggested in the case of NGC
4051 \citep{pounds11}.

The ubiquity of the soft excess means that any X-ray spectral fitting
of type 1 AGN must take this feature into account.  It has been shown
that leaving this component unmodelled can lead to a \gmm value
$\sim0.1$ too steep.  In addition, any attempt to constrain an
intrinsic \nh value must also include the blackbody component in the
fit since they appear in the spectra at a similar energy range.  It
was found in \scott that the average \gmm values for sources fit with
the `po+bb' model were significantly flatter than those fit with the
`po' model ($\Gamma\sb{\rm po}=1.98 \pm 0.01$, $\Gamma\sb{\rm
po+bb}=1.87 \pm 0.05$, KS significance = 0.0003) suggesting that the
underlying power-law slope in sources with a soft excess is different.
However, our modelling uses a blackbody at low energies to model the
soft excess component and this is purely phenomenological - it
provides a good representation of the feature seen in spectra of our
quality.  If the soft excess is actually a broad spectral feature as
suggested by reflection models, our modelling may not be fully
accounting for the spectral complexity and the power-law slope at
higher energies could still contain some of this component.

This work and many in the literature, have confirmed the presence of a
population of AGN which are classified as type 1 due to the presence
of broad emission lines in their UV/optical spectra, but also show
significant X-ray absorption.  Conversely there are also objects
optically classified as type 2 which are unabsorbed in X-rays
(e.g. \citealt{panessa02}, \citealt{mateos05a,mateos05b}).
These objects are not reconciled by the standard orientation based
Unified Model which invokes an obscuring torus to both block the line
of sight to the broad line region and give X-ray absorption in the
case of type 2 AGN, but not type 1.  Constraining the fraction of
absorbed type 1 sources may aid in interpreting these objects in terms
of a correction to the Unified Model.  Such a correction may include
invoking the `clumpy torus' model \citep{nenkova08a,nenkova08b} in
which the torus consists of individual clouds.  Observations of AGN
which showed large amplitude and rapid variability of the X-ray column
density were interpreted as occultations by clouds in the broad-line
region, suggesting clumpy absorbers may be present at a range of
different scales.  In this model observing a given source as absorbed
depends upon the covering factor of the clouds and is a probability
issue rather than one of just orientation \citep{risaliti02}.  It has
also been suggested that obscuration of AGN could occur due to the
presence of a warped accretion disc
\citep{greenhill03,nayakshin05,lawrence10}.  In this scenario there is
no need for an obscuring torus; the type 2 objects tend to be the ones
with larger misalignments, giving larger covering factors.

Whilst it would be interesting to compare how the detection of the low
energy spectral components varies with radio loudness\footnote{The
determination of the radio loudness of the sources is described in
S11.}, we lack the statistics to do this.  However we do note that
6/75 radio loud quasars (RLQ) have a detected soft excess and a joint
fit of 29 sources with a total of $\sim7000$ counts gives similar soft
excess parameters to those found for the radio-quiet objects. The
prevalence and magnitude of soft-excesses thus appears very similar
for RQQ and RLQ, contrary to previous suggestions that RLQ do not
include this component, (e.g. \citealt{sulentic10}).


\section{Conclusions}
\label{section:conclusions}
In this paper we have simulated X-ray spectra consisting of a power
law and a soft excess, in which the shape of the later component
represents a typical example of those found in the \citet{scott11}
sample.  The spectra were fit in \xspec over the energy range
0.5--12.0 keV and an F-test was used at 99\% significance to test
whether the component was statistically required in the fit.  By
repeating this procedure for spectra with different numbers of counts,
maximum detection curves were generated and compared to the observed
results from the real data sample presented in S11.  The effect of
redshift on the detectability was also taken into account.  Despite
the raw percentage of sources with a soft excess being $\sim8\%$, we
showed that after correcting for the spectral quality, the intrinsic
percentage is $75\pm23\%$.  This suggests that within the \scott
sample, almost all of the sources could include a soft excess
component with a shape and size typical of that seen in the highest
count spectra, and it is merely the quality of the spectra that is
limiting our ability to detect them.  The detectability of a soft
excess component is dependent on both the blackbody temperature and
the size of its normalisation with respect to the underlying power law
used in the modelling, but we find that using slightly different
values for either does not change our overall conclusion.

If soft excesses are ubiquitous, then the feature should be recovered
in a combination of low count spectra.  Groups of $\sim 50$ spectra
($\sim 7000$ counts) were created in narrow $z$ bins, including
spectra with $<500$ counts and which had no previous evidence for
additional spectral features.  The groups at $z<1$ were shown to be
better fit with a model including a soft excess, and the temperature
and normalisation with respect to the underlying power law of the
components required were consistent with those found in individual
high count spectra.

We are unable to conduct a simulation procedure in order to determine
the percentage of type 1 AGN which require an intrinsic cold
absorption component.  However, we suggest that its detectability may
not be as dependent upon spectral quality as the soft excess.  We
stress that a non-negligible percentage, $\sim 5\%$, of type 1 AGN may
include such an absorption feature and therefore any spectral
modelling must take the possibility of this feature into account.


\section*{Acknowledgments}
We thank the referee for comments which helped improve this paper.
AES acknowledges support from an STFC studentship and SM acknowledges
financial support from the Ministerio de Ciencia e Innovaci\'{o}n
under project AYA2009-08059 and AYA2010-21490-C02-01.  This work was
based on observations obtained with \textit{XMM-Newton}, an ESA
science mission with instruments and contributions directly funded by
ESA Member States and NASA.


\bibliographystyle{mn2e}
\bibliography{simbib}


\bsp

\label{lastpage}

\end{document}